\documentclass[runningheads]{llncs}
\usepackage[T1]{fontenc}
\usepackage{hyperref}
\usepackage{amsmath}
\usepackage{booktabs}
\usepackage{multirow}
\usepackage{makecell}
\usepackage{siunitx}
\usepackage{threeparttable}

\usepackage{graphicx}
\usepackage{float}
\usepackage{graphicx}
\usepackage{subcaption}
\usepackage{booktabs}
\usepackage{makecell}
\usepackage{xcolor}
\usepackage{colortbl}
\usepackage{bbding}
\usepackage{framed}
\usepackage{multirow}

\definecolor{PrakColor}{RGB}{186,223,121}       
\definecolor{G4lpcolor}{RGB}{248,214,232}    
\definecolor{S8ccolor}{RGB}{194,190,214}       
\definecolor{C8color}{RGB}{242,180,109}       
\definecolor{G8color}{RGB}{147,185,209}    
\definecolor{S8scolor}{RGB}{150,209,198}       
\definecolor{V8color}{RGB}{243,137,123}

\begin{document}

\title{Can Agents Win the Video Browser Showdown?}

%
%
\author{Bastian Jäckl\inst{1}\Envelope \orcidID{0009-0004-3341-1524} \and
Zuzana Vopálková\inst{1}\orcidID{0009-0000-7823-9511} \and Daniel A. Keim\inst{1}\orcidID{0000-0001-7966-9740} \and
Jakub Loko\v{c}\inst{2}\orcidID{0000-0002-3558-4144}}

\institute{University of Konstanz, Konstanz, Germany \and
Charles University, Prague, Czechia \\
\email{bastian.jaeckl@uni-konstanz.de}}
\authorrunning{Jäckl et al.}

\maketitle              
\begin{abstract}
Searching large video collections is typically an interactive process in which users play two roles. First, they hold the search intent: the underlying goal that determines what content they seek and why. Second, users must operationalize this intent through an iterative search loop. Users translate their intent into queries, browse the retrieved candidates, and refine their queries based on the results. In this paper, we investigate the capabilities of modern Vision Language Models (VLM) and agentic approaches to reach search goals interactively and fully autonomously. Specifically, we study whether a provided initial specification of a search goal might be sufficient to solve traditionally interactive search tasks with an agentic system. Provided that the involved VLMs are not aware of the whole large video dataset in advance, the key challenge lies in the effective combination of an existing interactive video search system and a smart VLM agent controlling the system. While the search system provides indexing and efficient querying, the VLM-based agents analyze top-ranked items and make decisions about next actions. Our results show that modern agents can autonomously operate interactive video retrieval systems to solve many search tasks from an initial intent description, achieving performance competitive with strong historical expert-operated systems in several settings.

\end{abstract}

\section{Introduction}
Modern video retrieval systems support interactive search through a wide range of capabilities, including textual, visual, and temporal queries, relevance feedback, filtering, keyframe layout algorithms, and video players~\cite{vbs_eval_2024}. Users must understand these capabilities and employ them effectively throughout the search process. Previous work has shown that operating such systems can be cognitively demanding, time-consuming, prone to errors, and reliant on expert knowledge~\cite{vbs_eval_2024,keyframeLayouts,vbs_eval_2020}. Consequently, successful retrieval may depend not only on the capabilities of the system but also on the user's experience.

Recent advances in AI agents suggest that parts of the interactive search process can be automated. Agents can interpret natural-language instructions, plan and execute tool calls, and inspect textual and visual information~\cite{yao2023reactsynergizingreasoningacting,wu}. These capabilities have enabled agents to understand long-form video content, answer questions about individual videos, and perform initial forms of autonomous video retrieval~\cite{fan2024videoagentmemoryaugmentedmultimodalagent,yuan2025videoexplorerthinkvideosagentic,zhang2024omagentmultimodalagentframework}. In particular, Wu et al.~\cite{wu} demonstrated impressive performance of an agentic approach to ad-hoc video search in which agents formulate textual queries, evaluate retrieved candidates, and decide whether to reformulate the query or continue exploring the current result set. However, it remains unclear whether agents can autonomously manage the broader range of search and interaction capabilities required across diverse interactive video search tasks. Moreover, agentic video search has not yet been systematically evaluated relative to strong expert-operated systems across such tasks.

To address these gaps, we establish an agentic performance baseline for interactive video search and contextualize it through comparison with expert human searchers. We isolate search execution from intent specification by treating a well-specified search intent as fixed input and investigating whether an agent can assume the operational role of a human searcher without further user intervention. Because real users may communicate their needs incompletely or ambiguously, this controlled setting provides an optimistic estimate of real-world performance. Specifically, we ask: \emph{Given a well-specified search intent, how effectively can an autonomous agent perform the operational role of a human searcher by formulating queries, browsing and assessing results, refining its search strategy, and determining whether retrieved candidates satisfy that intent, and how does its performance compare with that of expert human searchers?}

We investigate this question using the Video Browser Showdown (VBS) as our evaluation setting~\cite{vbs_eval_2024,vbs_eval_2023,vbs_eval_2021,vbs_eval_2020}. VBS is particularly suited for this purpose: its controlled task specifications provide fixed, well-specified search intents; its diverse task types and official time constraints support standardized evaluation; and its historical expert runs provide a human-performance reference. To this end, we develop a fully autonomous multi-agent video search system connected to the PraK retrieval backend~\cite{prak}. The system can issue textual, visual, and temporal queries, apply filters and relevance updates, and inspect individual keyframes and videos. A shared knowledge base records observations and intermediate results, enabling the agents to reason over collected evidence and plan multi-step search strategies. Together, the agents coordinate the complete operational search process across four VBS task types: textual known-item search, visual known-item search, ad-hoc video search, and visual question answering. We evaluate the framework on three years of historical VBS tasks, thereby establishing an agentic performance baseline and positioning it relative to corresponding expert human performance. The \textbf{main contributions} are:

\begin{itemize}
\item A fully autonomous multi-agent framework executing the complete search process across four VBS task types.
\item An agentic performance baseline on three years of controlled, well-specified VBS tasks, contextualized against strong historical expert runs.
\item Analyses of agent behavior and failure modes, with open-source code\footnote{\url{https://github.com/mmmAgenticVR/agenticvr}} and a visual interface for inspecting search traces.
\end{itemize}

\section{Related Work}
Our work is embedded between interactive video retrieval and agentic video search. Below, we summarize the state-of-the-art and position our work.

\textbf{Interactive Video Retrieval --} Interactive video retrieval systems generally comprise three components: (1) a preprocessed video collection represented through index structures, often based on keyframes; (2) retrieval mechanisms supporting different query modalities and refinement operations; and (3) a user interface for formulating queries, inspecting results, and browsing candidate videos. Video retrieval benchmarks such as the VBS~\cite{vbs_eval_2020,vbs_eval_2021,vbs_eval_2023,vbs_eval_2024,extended_evaluation_2022}, the Lifelog Search Challenge~\cite{lsc2021,lsc2022_2024}, and CASTLE~\cite{CASTLE} document an increasing reliance on powerful multimodal embedding models such as CLIP~\cite{radford2021clip} and SigLIP~\cite{zhai2023siglip}. These models enable semantic video retrieval through natural-language queries, allowing many search tasks to be addressed effectively through textual descriptions of the information need~\cite{loggingPaper}. However, tasks involving complex semantics, ambiguous descriptions, or visually similar candidates still require users to browse results, assess candidates, and iteratively refine their queries. Recent systems have begun to use LLMs to assist users by generating plans for executing such searches~\cite{autonomy_video_retrieval,lifeExplore2026,visioneLSC}. Whereas these approaches leave the execution of the search process and the assessment of candidates to the user, our work investigates whether an autonomous agent can perform the complete operational search process.

\textbf{Agentic Video Search --} Recent open-source VLMs support foundational video-understanding tasks such as captioning and question answering, while agentic workflows extend these capabilities to long-form video understanding \cite{qwen35blog,llava1,llava2,fan2024videoagentmemoryaugmentedmultimodalagent,yuan2025videoexplorerthinkvideosagentic,zhang2024omagentmultimodalagentframework}. In interactive retrieval, conversational approaches clarify intent or coordinate multimodal retrieval modules under continued user feedback~\cite{vired,liang,umivr,vinh2027conversational,visioneLSC,lifeExplore2026}. Other systems orchestrate query generation, planning, refinement, reranking, and validation through editable or predefined workflows \cite{mmmagents,autonomy_video_retrieval,maven,quan}. These approaches either retain user involvement or rely on predefined execution logic. More recent work closes the retrieval loop by adapting to intermediate results. VRAgent~\cite{vragent} iteratively refines multimodal retrieval instructions based on top-result confidence, while Wu et al.~\cite{wu} use VLM-based candidate assessment and stored reasoning traces to decide whether to continue inspecting results or reformulate the query.

Overall, existing approaches either retain the user in the loop, rely on predefined workflows, or focus primarily on textual ad-hoc retrieval. Our work extends closed-loop agentic search to four heterogeneous VBS task types and evaluates fully autonomous operation under official time constraints, including comparison with historical expert-operated systems.

\section{Agentic Video Search: Benchmark and System}\label{sec:system}
We establish an agentic performance baseline for interactive video retrieval by executing a fully autonomous system on historical VBS tasks and assessing its performance relative to expert runs. We first define the benchmark setting and evaluation protocol before describing our autonomous search system.

\subsection{VBS Setting and Tasks}
VBS evaluates time-constrained interactive search in large video collections through tasks that specify an information need using text, visual material, or both. Its controlled specifications, diverse task types, official time limits, and archived expert results provide a standardized benchmark for agentic search. We treat the task material as a fixed search intent from which the agents autonomously formulate queries, inspect candidates, refine their search, and produce a result under the same time limits and without human intervention.

\textbf{VBS Tasks --} We consider four VBS task types. In \emph{Textual Known-Item Search (t-KIS)}, participants receive a textual scene description and must identify the matching scene. In \emph{Visual Known-Item Search (v-KIS)}, they receive a video clip and must locate the corresponding scene in the collection. \emph{Ad-Hoc Video Search (AVS)} requires retrieving as many relevant scenes as possible for a textual topic or event description. Finally, in \emph{Video Question Answering (VQA)}, participants receive a video clip and a question and must locate the corresponding video and extract the requested information.

While t-KIS and AVS specify the search intent textually, v-KIS and VQA present visual material directly to human participants. Humans can repeatedly inspect this material, retain relevant perceptual cues, and translate them into queries and other search actions. However, VBS rules prohibit using task-provided videos or frames directly as retrieval queries, as this would bypass the intended search process and largely reduce the task to visual-similarity search. This creates a task-representation challenge: providing agents with only a textual description of the visual material may omit perceptual cues available to humans, whereas providing the original video or frames could enable direct or indirect visual-similarity search. No representation therefore yields a strictly equivalent comparison. We address this challenge through the \emph{Task Encoder}, which defines the task representation provided to the agents.

\textbf{Task Protocol and Task Encoder --}
We represent a VBS task as $\tau_m=(y,E_m(x),T)$, where $y$ is the task type, $x$ the task material, $T$ the time limit, and $E_m$ the Task Encoder mapping the material to the agent input under condition.

For textual tasks, $E_m$ preserves the original description and releases t-KIS hints according to the official schedule. For visual tasks, we exclude the original video and frames from the agent input and convert the visual material into a fixed textual description. We evaluate two encoder conditions $m$: a \emph{VLM Description}, generated automatically, and a \emph{Human Description}, used to distinguish limitations of automatic encoding from those of the subsequent search process. For VQA, the original question is appended to the respective description. The visual-task specification is provided before the timed search, after which agents cannot access the original material; all subsequent system components remain unchanged. Visual-task results therefore evaluate autonomous search from a textualized visual intent rather than under perceptual access equivalent to that of human participants.

\subsection{Agentic Search System}
Given the task specification $\tau_m$, our system executes a closed-loop retrieval process in which a central planner coordinates specialized agents and retrieval operations through a shared search state. We describe the modules below.

\begin{figure}
    \centering
    \includegraphics[width=\linewidth]{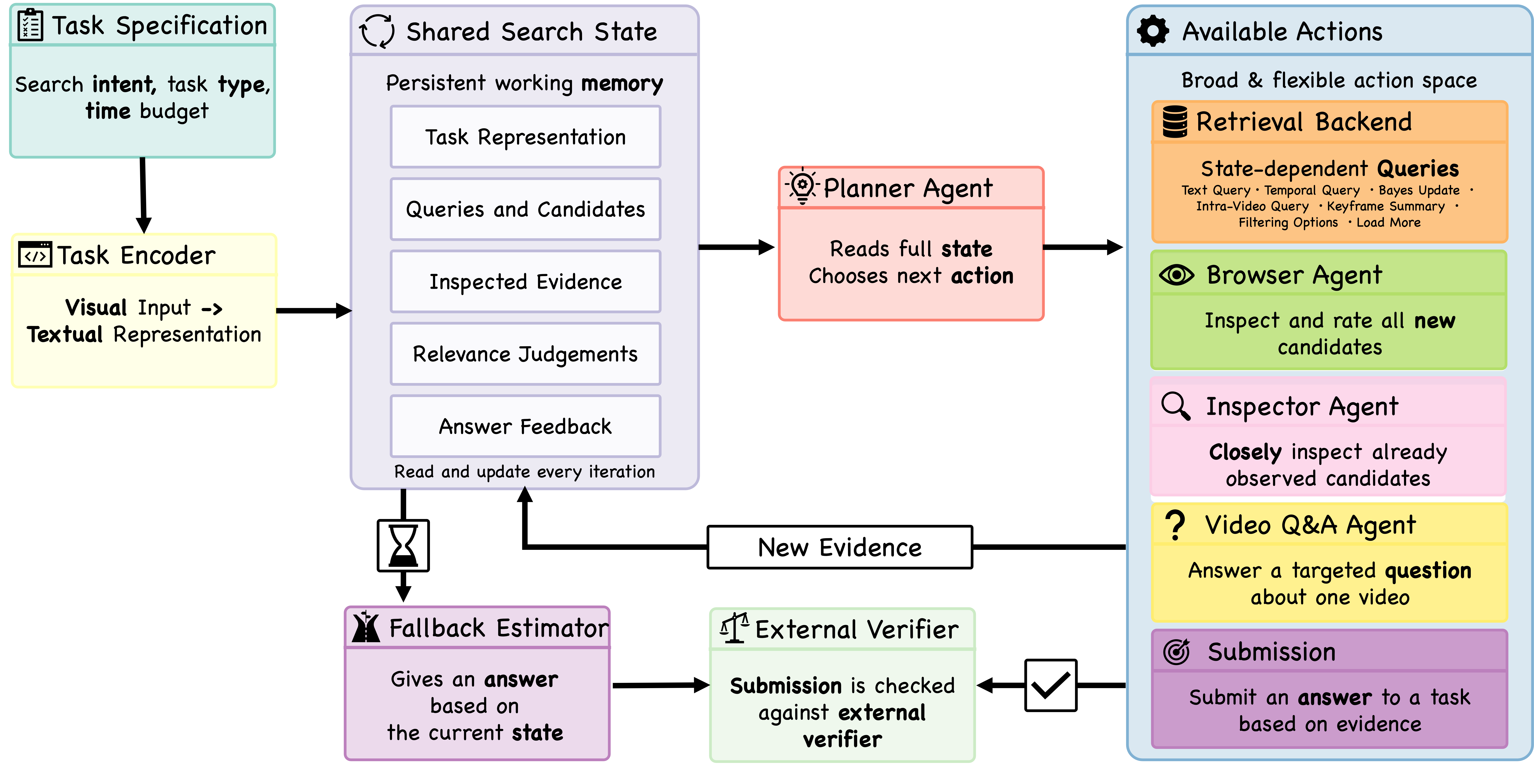}
    \caption{Workflow of the agentic retrieval loop.}
    \label{fig:placeholder}
\end{figure}

\textbf{Shared Search State --} At decision step $t$, all information accumulated during the search is stored in the persistent shared state
\[
s_t=\left(\tau_m,H_t,C_t,E_t,F_t\right),
\]
where $\tau_m$ is the encoded task specification under condition $m$, $H_t$ records executed actions, queries, and returned candidates, $C_t$ contains candidate descriptions and relevance assessments, $E_t$ stores additional video-level evidence, and $F_t$ records submissions and benchmark feedback provided by the external verifier.

The state serves as the system's memory. The planner reads it before selecting an action, while the executed retrieval and inspection modules add new observations. This enables the system to accumulate evidence, revisit previous results, account for rejected submissions, and avoid repeating unsuccessful actions.

\textbf{Planner Agent --} The planner agent orchestrates the retrieval process. At each step $t$, a VLM receives the current search state $s_t$ and a fixed planner instruction prompt $p$. The prompt is composed of structured natural-language specification of the planner's role, the available actions and their schemas, the applicable task constraints, and guidelines for selecting actions.

Conditioned on the current state $s_t$ and planner prompt $p$, the planner produces a structured action $a_t$ containing an action type, rationale, and arguments. Depending on its type, the validated action is routed to a retrieval tool, an inspection agent, or a terminal handler. Retrieval arguments specify the complete query, including text, filters, and referenced identifiers. Execution returns an observation $o_t$, such as retrieved candidates, inspection findings, or task feedback, which is incorporated into the next state:

$$
(s_t,p)
\xrightarrow{\operatorname{Planner}} a_t
\xrightarrow{\operatorname{Execute}} o_t
\xrightarrow{\operatorname{Update}} s_{t+1}.
$$


\textbf{Retrieval Backend --} Retrieval actions query the publicly available VBS 2026-winning PraK backend~\cite{prak,prakGithub} and add ranked keyframes to the shared state. PraK supports multimodal \textit{semantic search}, multi-query \textit{temporal search}, and memoryless \textit{Bayesian relevance feedback} using keyframes from the latest result set. \textit{Image-similarity search} uses a previously retrieved collection keyframe as its query. \textit{Intra-video} and \textit{localized search} restrict retrieval to a selected video or spatial region, while \textit{video-summary} loads a keyframe overview of a video.

The admissible action set $\mathcal{A}(s_t)$ depends on the current state. Preconditions are included in the planner prompt and enforced by a deterministic validation harness: relevance feedback may use only the latest result set, and identifier-based actions may reference only items in $s_t$. Retrieval actions additionally support \emph{one-per-video} filtering for diversity and \emph{unseen-only} filtering to exclude previously returned keyframes.

\textbf{Browser --} The browser agent is automatically invoked whenever a retrieval action returns new keyframes. It receives the retrieved candidates together with the encoded task specification $\tau_m$ and evaluates them in two stages. First, the agent inspects each keyframe and assigns a visual relevance label: \emph{irrelevant}, \emph{relevant}, or \emph{match}. Candidates labeled as relevant or matching are then examined in their surrounding temporal context. Based on this scene-level inspection, the agent assigns a second label using the same three categories. If the scene matches the task, the browser initiates an immediate submission. Otherwise, relevant observations are summarized and retained as evidence for subsequent planner decisions. The generated descriptions and relevance judgments are stored in the shared search state. Inspection actions and results update $H_t$, candidate assessments update $C_t$, extracted scene-level evidence updates $E_t$, and any submission and corresponding feedback update $F_t$.

\textbf{Inspector --} The planner may invoke the inspector to answer targeted questions about selected keyframes or videos. The resulting evidence complements the browser's automatic candidate assessment and is stored in $E_t$.

\textbf{Video Q\&A --} For VQA tasks, the VQA agent is invoked once the planner or browser identifies a candidate target video. Given the question, it selects one of three strategies. The \textit{relevance-first} strategy inspects the scenes surrounding the top-$k$ keyframes most relevant to $q$ and submits an answer once sufficient evidence is found; otherwise, it falls back to the \textit{exhaustive} strategy. This strategy collects evidence from all scenes before answering and is suited to global questions such as counting occurrences, but is computationally expensive. The \textit{chronological} strategy instead inspects scenes in temporal order until the answer is found, making it suitable for questions about the first occurrence of an event.

\textbf{Fallback Estimator --} Shortly before the time limit, the fallback estimator selects the most likely answer from the accumulated search state $s_t$ and submits it directly. This ensures that a final prediction is returned even when the retrieval loop has not identified a definitive answer.

\textbf{External Verifier --} The external verifier is not part of the retrieval loop but simulates the VBS submission system, such as DRES~\cite{dres}. For t-KIS and v-KIS, it accepts or rejects each submission. Rejections are recorded in $F_t$, allowing the planner to resume the search while avoiding previously rejected candidates. AVS and VQA submissions receive no online feedback and are evaluated only in post-processing; their retrieval loop therefore continues until the time limit.

\section{Experiments and Results}
\subsection{Implementation and Experimental Settings}

\textbf{VBS Tasks and Historical Expert Reference --} We evaluate all 81 tasks from the 2024--2026 VBS editions across the V3C~\cite{RossettoSAB19}, MVK~\cite{MVK} and LHE~\cite{medical} datasets. We contextualize agent performance using the three highest-performing \emph{expert} runs from each edition: \texttt{vibro1}, \texttt{visione1}, and \texttt{prak2} (2024); \texttt{nii-uit2}, \texttt{nii-uit1}, and \texttt{prak6} (2025); and \texttt{prak4}, \texttt{prak2}, and \texttt{prak1} (2026). We use official results~\cite{vbs_archive} and exclude novices.

\textbf{Implementation of the Agentic Search System --} We implement the agentic loop in \texttt{Python}. All agents use \texttt{Qwen3.8-27B-FP8}~\cite{qwen38_27b_fp8}, served locally through \texttt{vLLM} on a single NVIDIA H100 GPU, without proprietary model APIs. We disable the model's reasoning mode to reduce inference latency. The planner uses a temperature of 0.5, while all other agents use greedy decoding. 

Importantly, the autonomous agents use the same PraK retrieval backend that was operated by the PraK expert users in VBS 2026~\cite{prak}, using the same embedding model as already used in 2025. Thus, comparisons against the 2026 PraK runs largely control for the retrieval backend and primarily differ in who operates the system: an autonomous agent or an expert human user. Comparisons against runs from other systems, and against PraK configurations from earlier editions, remain subject to differences in retrieval-system configuration.


\textbf{Human Annotations --}
For the 51 tasks containing reference videos, six computer-science graduates
independently produced two textual descriptions per task (102 total) under a balanced assignment. Each scene was repeatedly played for three minutes without pausing,
approximating the VBS viewing conditions.

\textbf{Inference Protocol --} Task encoding is completed before the task timer starts and is excluded from the reported completion times. Accordingly, neither the three-minute human annotation period nor the generation of the VLM description is counted as search time. We evaluate retrieval depths of \(k\in\{4,16,50,100\}\) candidates per query. Larger values of $k$ increase per-query coverage but leave less time for additional retrieval iterations. For KIS and VQA, we selected $k$ by mean completion time, counting unsolved tasks at timeout; for AVS, by the number of correct submissions. At the task-specific timeout, the latest prediction of the fallback estimator is used as the final result. Further hyperparameters are provided with the code. We conduct three independent runs per configuration and report their individual results.

We emphasize that the comparison with real users, particularly for visual tasks, is only indicative. While human participants used different retrieval systems and could restart the search process from a different perspective, agents had the advantage of receiving the task encoding from the first second of the search process. A more grounded comparison between humans and agents remains an important direction for future work. Our results should therefore be interpreted as an agentic baseline contextualized by historical expert performance.

\subsection{Performance of Agentic Search}\label{sec:performance}
We compare agentic and historical expert performance in terms of retrieval success, completion time, and error rates.

\begin{figure}
    \centering
\includegraphics[width=\linewidth]{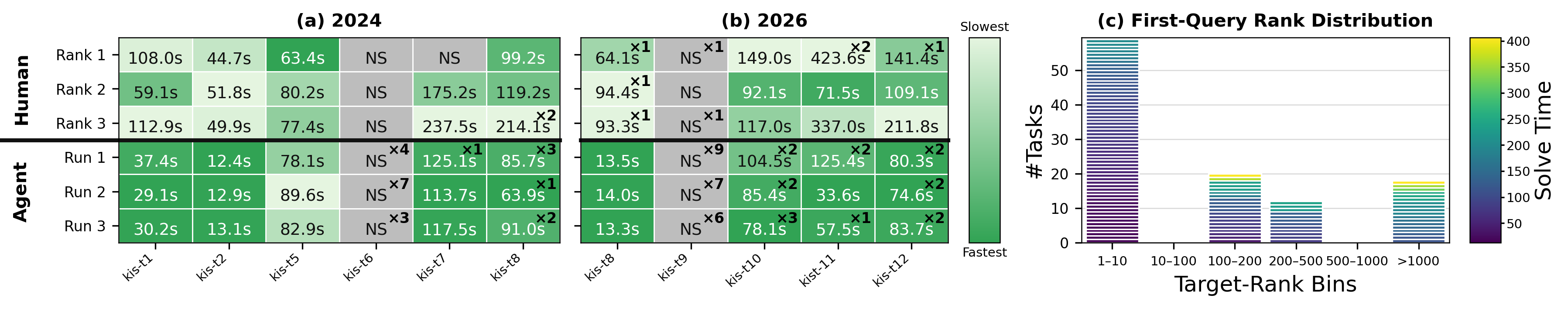}
    \caption{\textbf{a) and b):} t-KIS performance of agents ($k=4$) and humans. Color encodes solve time. Crosses at the top right show number of wrong answers. \textit{NS} means not solved. \textbf{c):} First-query target-rank distribution across all t-KIS runs ($k=4, k=16, k=50, k=100$). Each rectangle represents one successfully solved task, grouped by rank bin and colored by solve time.}
    \label{fig:tkis-performance}
\end{figure}

\textbf{Textual Known-Item Search. --} \autoref{fig:tkis-performance} places the best-performing candidate configuration, \(k=4\), alongside the expert runs. Each of the three independent agent runs solved nine of the 11 tasks. On eight of the nine tasks solved by both agents and experts, at least one agent run was faster than the fastest expert run. In the two tasks not solved by any agent run, the target was never retrieved.

Alongside the strong completion rate, the system produced 59 incorrect submissions (compared to only 10 from the experts): 30 originated from the browser, 23 from the planner, and six from the fallback estimator. The browser also accounted for all 27 correct submissions, while the planner did not submit any correct answers. Note that for other values of $k$, the planner contributed a low number of correct answers. While fallback errors are expected because it must submit at timeout, planner submissions are voluntary and could benefit from more conservative confidence thresholds. In six of its 23 errors, the planner selected a different timestamp from a video already rejected after a browser submission, indicating that stronger use of rejection feedback could avoid repeated attempts. Error timing suggests a second improvement opportunity: 27 of the 59 incorrect submissions (45.8\%) occurred before the second hint, and 28 of the 30 browser errors (93.3\%) occurred before the final hint. When these 28 early browser errors were re-evaluated post hoc using all three hints, the browser rejected 26 (92.9\%). This suggests that delaying browser submissions until the final hint, or requiring stronger evidence earlier, could prevent most false submissions.

We further analyze the target rank induced by the first retrieval action, which was always a semantic search query. \autoref{fig:tkis-performance} (c) includes every successful run across all evaluated values of $k$ and shows the target's rank under this initial query together with the eventual completion time. While many targets are already highly ranked, a substantial fraction of successful runs start from a poor initial ranking: in nearly half (46\%), the target lies outside the top 100, and in several cases outside the top 1000. Importantly, many of these tasks are nevertheless solved quickly. This rules out a simple \emph{query once, then browse} explanation for the observed performance: given the time required to inspect candidates, sequentially traversing the initial ranking would not reach these targets within the observed completion times. Instead, the agent uses intermediate evidence to reformulate its query or switch retrieval strategy, producing a new ranking in which the target becomes accessible. Thus, the agent does not simply browse deeper into the initial ranking, but uses subsequent retrieval actions to recover targets that would otherwise remain out of reach.

\begin{figure}[H]
    \centering
    \includegraphics[width=\linewidth]{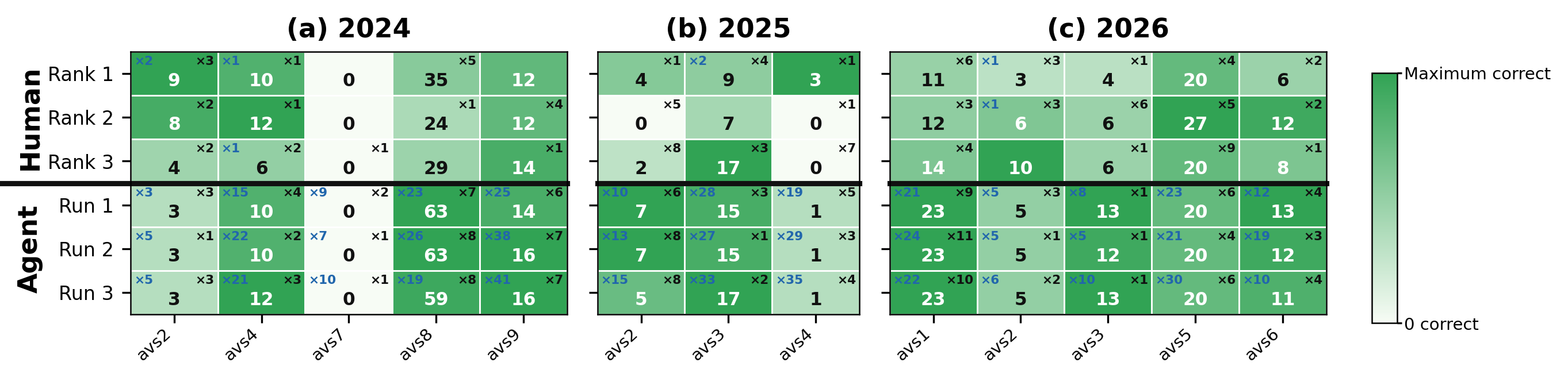}
    \caption{AVS-task performance with $k=100$ of the agents compared to humans. Color encodes the number of correct submissions. Black crosses show the number of wrong answers. Blue crosses show the number of unjudged answers. }
    \label{fig:avs-performance}
\end{figure}

\textbf{Ad-Hoc Video Search. --} \autoref{fig:avs-performance} reports AVS performance under the official VBS protocol, which credits at most one relevant submission per video. The archive contains judgments only for timestamps submitted during the original competitions and therefore does not provide exhaustive ground truth. To account for minor differences in timestamp selection, we transfer the judgment of the closest archived submission from the same video only if its timestamp differs by at most $\pm1$ second. If both positive and negative judgments fall within this window, we assign the label of the closest archived submission. All other submissions remain unjudged. The reported results therefore rely directly on the archived judgments with only minimal tolerance for timestamp variation, although unmatched submissions may still be relevant.

The agents retrieved relevant videos efficiently: Agents matched or exceeded the best expert run in eight of 13 tasks. Within a single 300-second run, the browser could inspect more than 300 unique scenes, demonstrating high candidate-assessment throughput. This helps explain both the strong retrieval performance and the high submission volume, with some individual runs producing more than ten judged-incorrect submissions. Manual inspection revealed that many errors were near-matches missing a specific task detail, identifying candidate verification as the primary bottleneck. We therefore evaluated a post-hoc falsification agent that searched each five-second candidate clip for evidence contradicting the task description. It rejected 56 of 169 judged-incorrect submissions (33.1\%), while falsely rejecting only 27 of 552 judged-relevant submissions (4.9\%). These results suggest that a second verification stage could substantially reduce errors while retaining most relevant results. In practice, humans could act as the second-stage reviewers.

\begin{figure}[H]
    \centering
    \includegraphics[width=\linewidth]{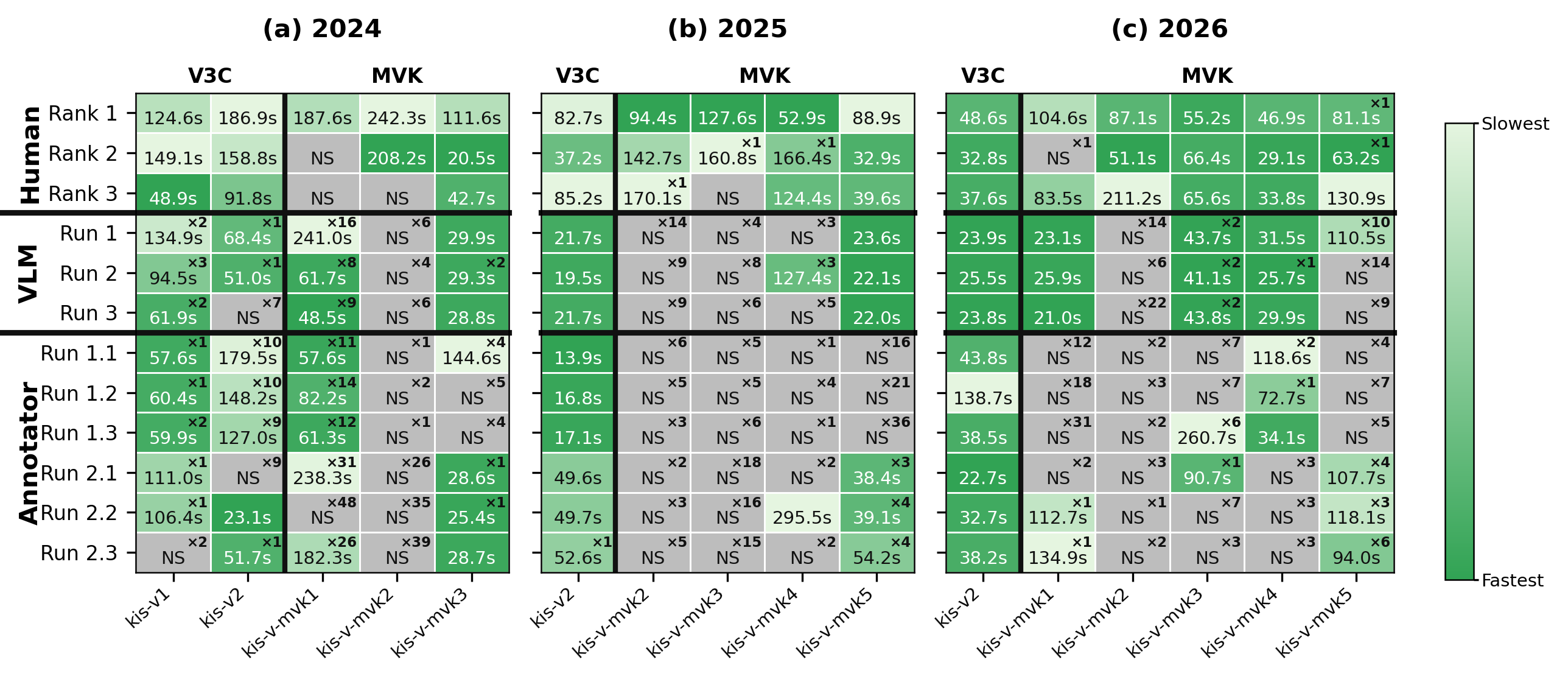}
    \caption{v-KIS performance of the agents compared to human performance using $k=16$. We show performance under two different task encodings: 1) \emph{VLM} is a description generated by a VLM  2) \emph{Annotator} is a description from humans.}
    \label{fig:vkis-performance}
\end{figure}

\textbf{Visual Known-Item Search. --}
\autoref{fig:vkis-performance} reports v-KIS performance on V3C and MVK. Agents approached expert performance on V3C but fell substantially behind on the more homogeneous MVK collection and solved none of the LHE tasks. Even with VLM-generated descriptions, all three agent runs failed on five of the 12 MVK tasks; with annotator descriptions, more than half of the individual runs were unsuccessful. These failures arose overwhelmingly during retrieval: in 146 of 154 unsuccessful MVK and LHE runs (94.8\%), the target never appeared among the candidate results, while it was retrieved but overlooked in only eight. The fixed textual task representation likely contributes to this gap. Humans can repeatedly inspect the target clip, perceive fine-grained cues, and compare them with retrieved candidates, whereas agents must rely on details preserved in the initial description. This asymmetry is particularly limiting in homogeneous collections and motivates updating the target representation during search.

\begin{figure}[H]
    \centering
    \includegraphics[width=\linewidth]{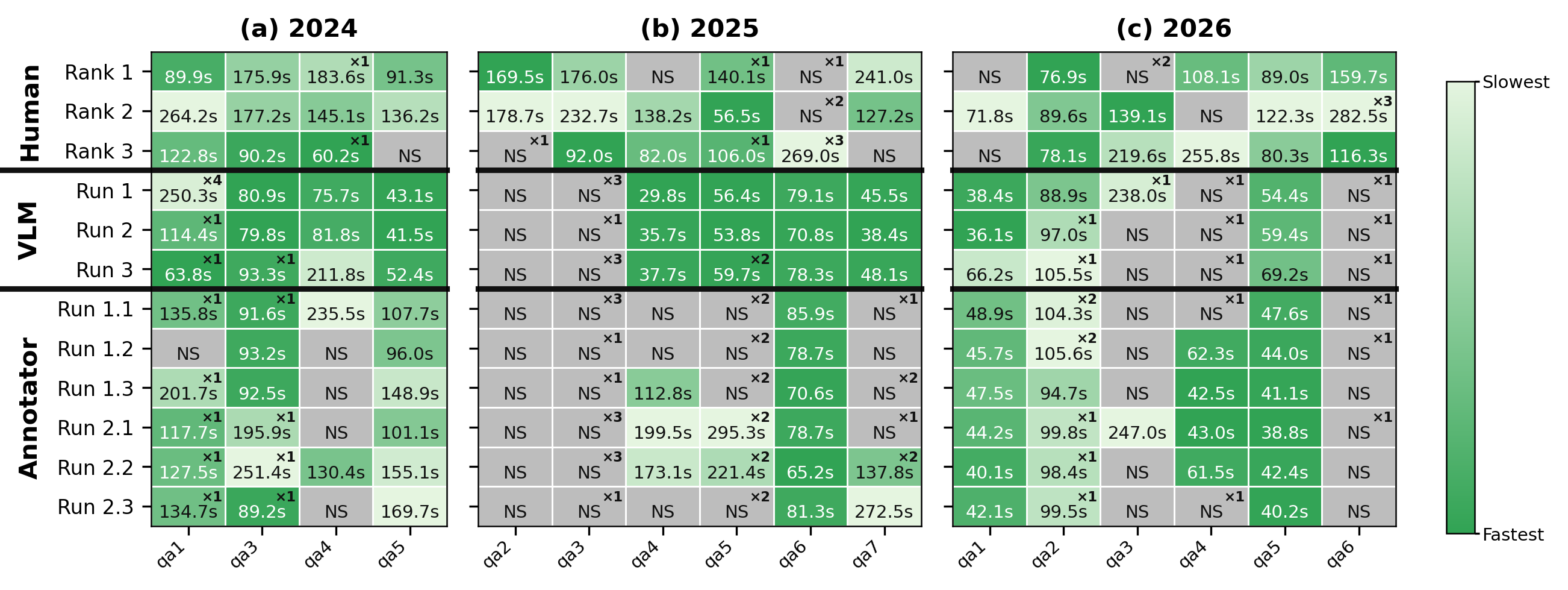}
    \caption{VQA performance of agents ($k=16$) and humans under both task encodings.}
    \label{fig:qa-performance}
    \vspace{-10mm}
\end{figure}

\textbf{Question Answering. --} For VQA, the three runs using VLM-generated task descriptions solved 11--12 of the 16 tasks, approaching the expert runs, which solved 12--14 tasks (\autoref{fig:qa-performance}). In each agent run, 7--8 of the successfully answered tasks were completed faster than all three experts. Four tasks remained unsolved across all VLM runs. In each case, the system retrieved the correct video and inspected the answer-containing scene but failed to extract the requested information. All four tasks required fine-grained object recognition, suggesting that stronger visual models or specialized perception modules could improve answer extraction. Human-annotated descriptions produced more variable results, with 7--12 solved tasks, as omitted details prevented retrieval of the target video.

\subsection{Behavioral Analysis}
\begin{figure}
    \centering
    \includegraphics[width=\linewidth]{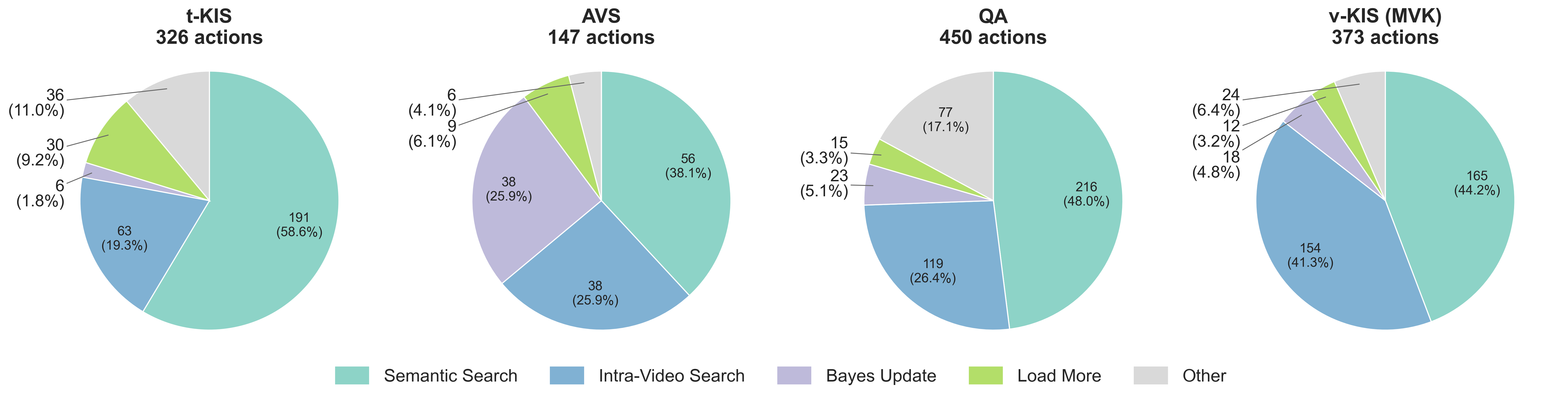}
    \caption{Number of actions per task predicted by the planner.}
    \label{fig:actions}
    \vspace{-5mm}
\end{figure}

\autoref{fig:actions} summarizes the planner's actions of the reported runs in \autoref{sec:performance}. Semantic Search (SemS) was the most frequent action in t-KIS (58.6\%), AVS (38.1\%), VQA (48.0\%) and v-KIS (44.2\%), consistent with human behavior~\cite{loggingPaper}. Intra-Video Queries (IVQ) were common (19.3--41.3\%) and nearly dominated v-KIS on MVK, while Bayes accounted for 25.9\% of AVS. Action bigrams show frequent query reformulation and alternation between collection and single-video search. The most frequent transitions were SemS \(\rightarrow\) IVQ (20.5\%), SemS \(\rightarrow\) SemS (13.3\%), IVQ \(\rightarrow\) IVQ (10.8\%), IVQ \(\rightarrow\) SemS (9.5\%), and SemS \(\rightarrow\) Bayes (6.5\%).

To assess whether the iterative retrieval loop responds to observed evidence rather than selecting follow-up actions independently of search outcomes, we examined the association between browser judgments after SemS and the planner's next refinement action using $k=16$ for all tasks. When the browser returned no relevant keyframe, the planner reformulated the query in 78.2\% of cases, compared with 18.4\% for IVQ and 3.4\% for Bayes. This strong association indicates that the planner uses negative retrieval feedback to adapt its strategy. Once relevant results were available, however, their number alone provided little separation: the preceding search returned a median of five relevant keyframes before reformulation, six before IVQ, and four before Bayes Update. Even with at least 10 relevant keyframes, IVQ (55.5\%) and SemS (40.4\%) remained the most frequent actions. Thus, an absence of relevant results strongly favors reformulation, whereas the choice among refinement strategies cannot be explained by result counts alone and is conditioned on the browser's qualitative descriptions together with the complete search state. This is consistent across varying $k$. Because these decisions also reflect the defined policy, this analysis characterizes responsiveness rather than optimality.

We further examined whether access to multiple retrieval actions contributed to success beyond the initial SemS action. After excluding runs in which the target was retrieved by the first action, 30.9\% of targets were reached through other retrieval strategies, predominantly IVQ (26.6\%). This demonstrates the value of giving the planner an action space beyond text-query formulation: alternative tools directly recovered targets not reached by the initial semantic search. A text-query-only agent remains a useful future baseline for isolating the value of adaptive tool selection and agent-specific retrieval policies.

\section{Conclusion}
We presented a fully autonomous agentic video search system and evaluated it on 81 historical VBS tasks, establishing an end-to-end baseline for autonomous interactive video retrieval. Given a well-specified search intent, the agents achieved performance comparable to strong historical expert-operated systems in several t-KIS, AVS, and VQA settings, while limitations remained for homogeneous visual collections and fine-grained information extraction. Behavioral analysis showed that the agents can formulate queries, execute different retrieval tools, assess candidates, and adapt their strategy based on intermediate feedback. The rapid recovery of targets ranked beyond the inspectable range after the initial query further demonstrates the importance of the iterative retrieval loop.

These results constitute an initial agentic baseline rather than a controlled human--agent comparison. They suggest a new division of labor: users specify intent, agents execute and adapt retrieval operations, and humans provide relevance feedback and final verification, particularly for fine-grained judgments. Future interfaces may therefore shift from manual tool operation toward accurate intent specification and oversight of autonomous search.

%
%
%
\begin{credits}
\subsubsection{\ackname} The authors acknowledge support by the state of Baden-Württemberg through bwHPC.
\end{credits}
\clearpage
\bibliographystyle{acm}
\bibliography{references}

@misc{qwen35blog,
    title = {Qwen3.5: Accelerating Productivity with Native Multimodal Agents},
    url = {https://qwen.ai/blog?id=qwen3.5},
    author = {Qwen},
    month = {February},
    year = {2026}}

@article{llava1,
  title   = {{LLaVA}-Video: Video Instruction Tuning With Synthetic Data},
  author  = {Zhang, Yuanhan and Wu, Jinming and Li, Wei and Li, Bo and
             Ma, Zejun and Liu, Ziwei and Li, Chunyuan},
  journal = {TMLR},
  year    = {2025},
  url     = {https://openreview.net/forum?id=EElFGvt39K}
}

@inproceedings{zhai2023siglip,
  title={Sigmoid loss for language image pre-training},
  author={Zhai, Xiaohua and Mustafa, Basil and Kolesnikov, Alexander and Beyer, Lucas},
  booktitle={ICCV},
  pages={11975--11986},
  year={2023},
  publisher={{IEEE}},
  doi={10.1109/ICCV51070.2023.01100.}
}

@inproceedings{vired,
author = {Madasu, Avinash and Oliva, Junier and Bertasius, Gedas},
title = {Learning to Retrieve Videos by Asking Questions},
year = {2022},
doi = {10.1145/3503161.3548361},
booktitle = {MM},
pages = {356–365},
numpages = {10},
}

@INPROCEEDINGS{liang,
  author={Liang, Kaiqu and Albanie, Samuel},
  booktitle={ICCV}, 
  title={Simple Baselines for Interactive Video Retrieval with Questions and Answers}, 
  year={2023},
  volume={},
  number={},
  pages={11057-11067},
  doi={10.1109/ICCV51070.2023.01018}}

@INPROCEEDINGS{umivr,
  author={Zhang, Bingqing and Cao, Zhuo and Du, Heming and Li, Yang and Li, Xue and Liu, Jiajun and Wang, Sen},
  booktitle={ICCV}, 
  title={Quantifying and Narrowing the Unknown: Interactive Text-to-Video Retrieval Via Uncertainty Minimization}, 
  year={2025},
  pages={22120-22130},}

@InProceedings{autonomy_video_retrieval,
author="Ho-Le, Minh-Quan
and Ho, Duy-Khang
and Ninh, Tu V.
and Gurrin, Cathal
and Tran, Minh-Triet",
title="From Expert Practices to Intelligent Agents: Autonomy in Interactive Video Retrieval",
booktitle="MMM",
year="2026",
doi="10.1007/978-981-95-6963-2_20",
pages="191--198",
}

@inproceedings{loggingPaper,
author = {J{\"a}ckl, Bastian and others},
title = {What Drove Success at the 15th Video Browser Showdown? A Comprehensive Interaction-Logging Analysis},
year = {2026},
doi = {10.1145/3805622.3810635},
booktitle = {ICMR},
pages = {1730–1739},
numpages = {10},
}

@inproceedings{radford2021clip,
  title={Learning transferable visual models from natural language supervision},
  author={Radford, Alec and others},
  booktitle={{ICML}},
  pages={8748--8763},
  year={2021},
}

@inproceedings{llava2,
  title     = {{SlowFast-LLaVA}-1.5: A Family of Token-Efficient Video
               Large Language Models for Long-Form Video Understanding},
  author    = {Xu, Mingze and others},
  booktitle = {COLM},
  year      = {2025},
}

@ARTICLE{vbs_eval_2024,
  author={Vadicamo, Lucia and others},
  journal={IEEE Access}, 
  title={Evaluating Performance and Trends in Interactive Video Retrieval: Insights From the 12th VBS Competition}, 
  year={2024},
  volume={12},
  number={},
  pages={79342-79366},
  doi={10.1109/ACCESS.2024.3405638}
}

@inproceedings{keyframeLayouts,
author = {J{\"a}ckl, Bastian and Kruchina, Ji{\v r}{\'i} and Joos, Lucas and Keim, Daniel A. and Peska, Ladislav and Lokoc, Jakub},
title = {Evaluating Keyframe Layouts for Visual Known-Item Search in Homogeneous Collections},
year = {2026},
booktitle = {ICMR},
pages = {1768–1777},
numpages = {10},
}

@article{lsc2022_2024,
  title={The State-of-the-Art in Lifelog Retrieval: A Review of Progress at the ACM Lifelog Search Challenge Workshop 2022-24},
  author={Allie Tran and others},
  journal={ArXiv},
  year={2025},
  numpages={24},
  doi={10.1109/ACCESS.2025.3644952}
}

@ARTICLE{lsc2021,
  author={Tran, Ly-Duyen and others},
  journal={{IEEE Access}}, 
  title={{Comparing Interactive Retrieval Approaches at the Lifelog Search Challenge 2021}}, 
  year={2023},
  volume={11},
  number={},
  pages={30982-30995},
  doi={10.1109/ACCESS.2023.3248284}}

@inproceedings{CASTLE,
author = {Rossetto, Luca and Bailer, Werner and Gurrin, Cathal and Dang Nguyen, Duc Tien and Schoeffmann, Klaus and Tran, Allie},
title = {{Overview of the First CASTLE Grand Challenge at ACM Multimedia 2025}},
year = {2025},
isbn = {9798400720352},
doi = {10.1145/3746027.3760242},
booktitle = {MM},
pages = {14271–14272},
numpages = {2},
}

@article{vbs_eval_2023,
author = {Loko\v{c}, Jakub and others},
title = {Interactive video retrieval in the age of effective joint embedding deep models: lessons from the 11th VBS},
year = {2023},
volume = {29},
number = {6},
doi = {10.1007/s00530-023-01143-5},
journal = {Multimedia Syst.},
pages = {3481–3504},
numpages = {24},

doi = {10.1007/s00530-023-01143-5}
}

@article{vbs_eval_2020,
author = {Loko\v{c}, Jakub and others},
title = {Is the Reign of Interactive Search Eternal? Findings from the Video Browser Showdown 2020},
year = {2021},
volume = {17},
number = {3},
doi = {10.1145/3445031},
journal = {TOMM},
articleno = {91},
numpages = {26},

}

@article{vbs_eval_2021,
  author  = {Heller, Silvan and others},
  title   = {Interactive video retrieval evaluation at a distance: comparing sixteen interactive video search systems in a remote setting at the 10th Video Browser Showdown},
  journal = {Int. J. Multimed. Inf. Retr.},
  year    = {2022},
  volume  = {11},
  number  = {1},
  pages   = {1--18},
  doi = {10.1007/s13735-021-00225-2}
}

@misc{qwen38_27b_fp8,
  author  = {{Qwen Team}},
  title   = {{Qwen3.8-27B-FP8}},
  howpublished = {\url{https://huggingface.co/Qwen/Qwen3.8-27B-FP8}},
  note         = {Accessed: 2026-08-29}
}

@misc{vbs_archive,
  author       = {Rossetto, Luca},
  title        = {{VBS-Archive}: Archive of Tasks and Results of the Video Browser Showdown},
  howpublished = {\url{https://github.com/lucaro/VBS-Archive}},
  note         = {Accessed: 2026-08-29}
}

@article{extended_evaluation_2022,
  title={Interactive multimodal video search: an extended post-evaluation for the VBS 2022 competition},
  author={Konstantin Schall and Werner Bailer and Kai Barthel and Fabio Carrara and Jakub Lokoc and Ladislav Peska and Klaus Schoeffmann and Lucia Vadicamo and Claudio Vairo},
  journal={Int. J. Multimed. Inf. Retr.},
  year={2024},
  volume={13},
  numpages={13},
  doi={10.1007/s13735-024-00325-9}
}

@inproceedings{
yao2023reactsynergizingreasoningacting,
title={ReAct: Synergizing Reasoning and Acting in Language Models},
author={Shunyu Yao and Jeffrey Zhao and Dian Yu and Nan Du and Izhak Shafran and Karthik R Narasimhan and Yuan Cao},
booktitle={ICLR},
year={2023},
url={https://openreview.net/forum?id=WE_vluYUL-X}
}

@InProceedings{prak,
author="J{\"a}ckl, Bastian
and others",
title="PraK V4 at the Video Browser Showdown 2026",
booktitle="MMM",
year="2026",
pages="230--237",
}

@INPROCEEDINGS{vragent,
  author={Shah, Ketul and Nathani, Pankaj and Chellappa, Rama and Heilbron, Fabian Caba},
  booktitle={WACV}, 
  title={VRAgent: Self-Refining Agent for Zero-Shot Multimodal Video Retrieval}, 
  year={2026},
  pages={8157-8167},
  doi={10.1109/WACV61042.2026.00787}}

@inbook{wu,
author = {Wu, Jiaxin and Wei, Xiao-Yong and Li, Qing},
title = {Adaptive Multi-Agent Reasoning for Text-to-Video Retrieval},
year = {2026},
booktitle = {ICMR},
pages = {1598–1607},
numpages = {10}
}

@InProceedings{maven,
author="Ngo, Quang Duc
and Nguyen, Hai Long
and Tran, Le Huy
and Vu, Tung Linh",
title="MAVEN: Video Retrieval System Using a Multi-agent Visual Exploration Network",
booktitle="SOICT",
year="2025",
pages="294--304",
}

@INPROCEEDINGS{quan,
  author={Quan, Khanh-An C. and Nguyen, Qui Ngoc and Luu, Duc-Tuan},
  booktitle={CVPRW}, 
  title={Toward Automation in Text-Based Video Retrieval with LLM Assistance}, 
  year={2025},
  volume={},
  number={},
  pages={3699-3707},
  doi={10.1109/CVPRW67362.2025.00355}}

@inproceedings{yuan2025videoexplorerthinkvideosagentic,
author = {Yuan, Huaying and Liu, Zheng and Zhou, Junjie and Qian, Hongjin and Shu, Yan and Sebe, Nicu and Wen, Ji-Rong and Dou, Zhicheng},
title = {VideoExplorer: Advancing Long-Horizon Video Understanding via Hierarchical Orchestration},
year = {2026},
isbn = {9798400722592},
booktitle = {KDD},
pages = {6350–6361},
numpages = {12},
}

@InProceedings{fan2024videoagentmemoryaugmentedmultimodalagent,
author="Fan, Yue
and others",
title="VideoAgent: A Memory-Augmented Multimodal Agent for Video Understanding",
booktitle="ECCV",
year="2025",
pages="75--92",
doi="10.1007/978-3-031-72670-5_5"
}

@inproceedings{zhang2024omagentmultimodalagentframework,
    title = "{O}m{A}gent: A Multi-modal Agent Framework for Complex Video Understanding with Task Divide-and-Conquer",
    author = "Zhang, Lu  and
      Zhao, Tiancheng  and
      Ying, Heting  and
      Ma, Yibo  and
      Lee, Kyusong",
    
    booktitle = "EMNLP",
    month = nov,
    year = "2024",
    
    doi = "10.18653/v1/2024.emnlp-main.559",
    pages = "10031--10045",
}

@InProceedings{mmmagents,
author="Ma, Zhixin
and Wu, Jiaxin
and Ngo, Chong Wah",
title="Leveraging LLMs and Generative Models for Interactive Known-Item Video Search",
booktitle="MMM",
year="2024",
pages="380--386",
}

@InProceedings{vinh2027conversational,
author="Vinh, Nguyen Mai
and others",
title="Towards Conversational Video Retrieval with an Intelligent Search Agent",
booktitle="SOICT",
year="2026",
}

@inproceedings{lifeExplore2026,
author = {Leopold, Mario and Tashtarian, Farzad and Schoeffmann, Klaus},
title = {lifeXplore 2026 - Lifelog Retrieval with Multilingual Vision-Language Encoders},
year = {2026},
booktitle = {LSC},
pages = {24–29},
numpages = {6},

}

@inproceedings{visioneLSC,
author = {Amato, Giuseppe and others},
title = {VISIONE: Redesigning an Interactive Retrieval System for Lifelog Search},
year = {2026},
booktitle = {LSC},
pages = {48–53},
numpages = {6},
}

@inproceedings{medical,
author = {Nasirihaghighi, Sahar and Ghamsarian, Negin and Peschek, Leonie and Munari, Matteo and Husslein, Heinrich and Sznitman, Raphael and Schoeffmann, Klaus},
title = {GynSurg: A Comprehensive Gynecology Laparoscopic Surgery Dataset},
year = {2025},
booktitle = {MM},
pages = {13141–13147},
numpages = {7}
}

@inproceedings{RossettoSAB19,
  author    = {Luca Rossetto and
               Heiko Schuldt and
               George Awad and
               Asad A. Butt},
  title     = {{V3C} - {A} Research Video Collection},
  booktitle = {MMM},
  pages     = {349--360},
  year      = {2019},
  doi       = {10.1007/978-3-030-05710-7\_29}
}

@inproceedings{MVK,
author="Truong, Quang-Trung
and others",
  title        = {Marine Video Kit: {A} New Marine Video Dataset for Content-Based Analysis
                  and Retrieval},
  booktitle    = {MMM},
  volume       = {13833},
  pages        = {539--550},
  address = {Berlin, Heidelberg},
  year         = {2023},
}

@article{dres,
author = {Sauter, Loris and Gasser, Ralph and Schuldt, Heiko and Bernstein, Abraham and Rossetto, Luca},
title = {Performance Evaluation in Multimedia Retrieval},
year = {2024},
doi = {10.1145/3678881},
journal = {TOMM},
month = oct
}

@misc{prakGithub,
   author  = {Stroh, Michael},
  title   = {{VideoRetrieval}: PraK Video Retrieval System},
  howpublished = {\url{https://github.com/IthronMinya/VideoRetrieval}},
  note         = {Accessed: 2026-08-29}
}

\end{document}